\documentclass[useAMS,usenatbib]{biom}

\usepackage[figuresright]{rotating}

\title[Assessing Impact of Unobserved Confounders with Sensitivity Index Probabilities]{Assessing Impact of Unobserved Confounders with Sensitivity Index Probabilities through Pseudo-Experiments}

\author{Beilin Jia$^{1}$, 
Donglin Zeng$^{1}$, 
Qing Yang$^{2}$,
Wei Pan$^{2,3,*}$\email{wei.pan@duke.edu; corresponding author.}\\
$^{1}$Department of Biostatistics, University of North Carolina at Chapel Hill\\
$^{2}$School of Nursing, Duke University\\
$^{3}$Department of Population Health Sciences, School of Medicine, Duke University}

\begin{document}


\date{{\it Received XXXX} 0000. {\it Revised XXXX} 0000.  {\it
Accepted XXXX} 0000.}



\pagerange{\pageref{firstpage}--\pageref{lastpage}} 
\volume{00}
\pubyear{0000}
\artmonth{XXXX}


\doi{00.0000/x.0000-0000.0000.00000.x}


\label{firstpage}


\begin{abstract}
Unobserved confounders are a long-standing issue in causal inference using propensity score methods. This study proposed nonparametric indices to quantify the impact of unobserved confounders through pseudo-experiments with an application to real-world data. The study finding suggests that the proposed indices can reflect the true impact of confounders. It is hoped that this study will lead to further discussion on this important issue and help move the science of causal inference forward. 
\end{abstract}

%

\begin{keywords}
Causal Inference; Propensity Score Methods; Sensitivity Analysis; Unobserved Confounders.
\end{keywords}


\maketitle


%

\section{Introduction}
\label{s:intro}

Causal inference is one of the most important goals of intervention studies in the biosciences. To estimate the causal effect of an intervention (or treatment), it is necessary to sufficiently control for potential confounders. There are many techniques for dealing with confounders so as to obtain unbiased causal estimates, such as restriction, standardization, stratification, matching, regression, and randomization \citep{Kahlert2017, Kestenbaum2009, Pourhoseingholi2012}. Among them, propensity score methods \citep{Rosenbaum1983b, Pan2015} are one of the most popular techniques. In propensity score methods, propensity scores associated with each treatment are first estimated by computing probabilities that subjects receive one particular treatment given all potential confounders. Then, propensity scores can be used to reduce selection bias from the confounders by balancing the distributions of the confounders between the treatment conditions through matching, stratification, weighting, or doubly robust estimation \citep{Pan2015, Pan2018}. Once subjects are balanced by their propensity scores, causal effects can be unbiasedly estimated by comparing responses of the balanced subjects to outcomes.

Unfortunately, propensity score methods can only deal with overt bias from observed confounders with the assumption that there are no unobserved (or unmeasured) confounders in the data. This assumption is called \textit{the strong ignorability in treatment assignment} by \citet{Rosenbaum1983b}, which is, however, statistically untestable. There have been several attempts to address the sensitivity to unobserved confounders specifically for propensity score methods, such as propensity score-based approach \citep{Li2011}, propensity score calibration \citep{Sturmer2005}, and high-dimensional propensity score adjustment \citep{Schneeweiss2009} among many other general techniques of sensitivity analysis that can be applied to propensity score methods for unobserved confounders \citep{Arah2008, Brumback2004, Carnegie2016, Fogarty2019, Groenwold2010, Hsu2013, Lin1998, McCandless2007, Rosenbaum1983a, Rosenbaum2009, Schneeweiss2006, VanderWeele2011, VanderWeele2017, Wang2006, Zhao2019}.

\begin{figure}[b]\centering
	\includegraphics[width=3in]{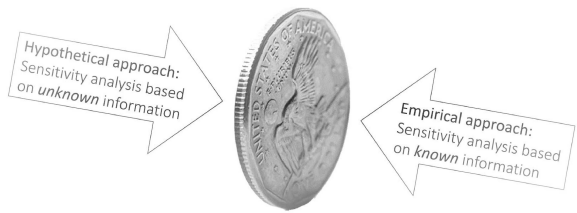} 
	\caption{Two approaches to sensitivity analysis: Two sides of the same coin}
	\label{figure:coin}
\end{figure}

Both specific and general techniques of sensitivity analysis for propensity score methods can help us understand how sensitive propensity score methods are to unobserved confounders. However, in those techniques, sensitivity is normally assessed based on \textit{unknown} information or a hypothetical range of the impact of unobserved confounders. Such \textit{hypothetical} approach to sensitivity analysis only gives us part of the picture about unobserved confounders, and the uncertainty among researchers about their research findings still remains \citep{Pan2003, Pan2016}. To better understand the impact of unobserved confounders on propensity score methods, it would be desirable to also assess the sensitivity based on \textit{known} information or the empirical evidence in the data. This \textit{empirical} approach to sensitivity analysis is on of the two sides of the same coin of sensitivity analysis with the hypothetical approach on the other side (Figure \ref{figure:coin}), but the former can help us understand how \textit{robust} propensity score methods are to unobserved confounders given the empirical evidence we have in the data so that researchers would have confidence in their research findings. The empirical approach to sensitivity analysis is particularly meaningful with the increased availability of large observational data, such as electronic health records, that possess rich information about almost all potential confounders that are observed either directly (i.e., observed confounders) or indirectly (related to unobserved confounders). 

Moreover, most of the current techniques of sensitivity analysis focus on the impact of unobserved confounders on causal effects estimated from outcome models with propensity score methods. Alternatively or more logically, as the first and utmost step of propensity score methods, propensity score estimation can also be the focal point. If propensity score estimates are robust to unobserved confounders, so are propensity score methods. As such, \citet{Pan2016} proposed a novel robustness index of propensity score estimation to unobserved confounders by quantifying the extremity of the propensity score for any given subject. The extremity is calculated as the tail probability of a parametric Pearson distribution \citep{Pearson1895} which has the same first four moments as the empirical distribution. Therefore, the insufficiency in observed confounders leads to high extremity probability. However, one limitation of this technique is that the extremity probability may be sensitive to the parametric distribution assumption. 

In this work, we adopted a similar idea to \citet{Pan2016} to study the extremity of certain index due to unobserved confounders. Because it is impossible to know unobserved confounders, we assume that the behaviors of unobserved confounders are similar to those of the observed ones distribution-wise. Therefore, we performed pseudo-experiments by utilizing the jackknife (or leave-one-out) technique that treats one of the observed confounders as an unobserved while the rest as observed confounders. We used the latter to obtain the empirical distribution of the influence scores--the change of propensity scores after excluding one particular variable. We then calculated the extremity of the influence score for the experimented unobserved confounder, as compared to the empirical distribution. We defined the influence score as \textit{sensitivity index probability} (SIP). 

Through pseudo-experiments, we obtained SIPs for all potentially unobserved confounders with respect to the ones among the observed confounders. There are several advantages of the proposed method. First, all SIPs are computed based on ranks, thus robust to parametric distribution assumptions. Second, the derived SIPs are useful for us to understand the impact of unobserved confounders if they look similar or are related to observed ones. Third, our simulated numerical evidence showed that the proposed SIPs are consistent of the true impact of the confounders when there is no unobserved one, but can deviate largely when there is. Motivated by the latter, in our application, we compared the derived SIPs to the ones under the null that the observed confounders are known to be sufficient to identify sufficiency or insufficiency of the confounders for a given subject.

\section{Method}
\label{s:method}
\subsection{Sensitivity index probability}
  
Consider any subset of confounders, $\{X_j, j\in {\cal A}\}$ where ${\cal A}$ is the index set. Following \citet{Pan2016}, we define a propensity score (PS) associated with ${\cal A}$ as
 $$e_{\cal A}(x)=\alpha_{\cal A}^*+x_{\cal A}^T\beta_{\cal A}^*,$$ 
 where $x_{\cal A}$ is the corresponding subset of $X=x$ and  $(\alpha_{\cal A}^*, \beta_{\cal A}^*)$ maximizes
$$l_{\cal A}(\alpha_{\cal A}, \beta_{\cal A})\equiv
E\left[Z (\alpha_{\cal A}+X_{\cal A}^T\beta_{\cal A})-\log (1+\exp\{\alpha_{\cal A}+X_{\cal A}^T\beta_{\cal A}\})\right],$$
where $Z$ is a binary treatment variable. Essentially, we perform a standard logistic regression by regressing treatment assignments on the variables in ${\cal A}$ to estimate $e_{\cal A}(x)$.

To examine the impact of one particular confounder on the propensity score, we adopt the leave-one-variable-out method. More specifically, for any given confounder in ${\cal A}$, say $X_j$, the influence score from  this confounder is defined as
$$\Delta_{j, {\cal A}}(x)\equiv e_{{\cal A}}(x)-e_{{\cal A}/j}(x).$$
In other words, the influence score for confounder $j$ with respect to set ${\cal A}$ is the magnitude change of the propensity score after deleting the $j$th variable from ${\cal A}$. Therefore, the larger $\Delta_{j, {\cal A}}(x)$ is, the higher impact of confounder $j$ has on the propensity score.

Our main goal is to evaluate whether the whole set of observed confounders, denoted as $\{X_1,...,X_K\}$,  are complete. In other words, whether there is any unobserved confounder that we should be concerned about. Because we can only rely on the observed data, we assume that any unobserved confounder can be approximated by one of the observed confounders. This assumption has been established in the literature on the similar topics \citep{Frank2000, McCandless2008, Pan2003, Pan2016}. With this assumption, we consider conducting the following $K$ pseudo-experiments. Let ${\cal N}=\{1,...,K\}$.  For any given $j=1,..,K$, we pretend that $X_j$ is an ``unobserved" confounder while the rest variables,  $\{X_k, k\in {\cal N}/j\}$, consist of all observed confounders. With the ``observed" confounders, we can obtain their influence scores, $\Delta_{k, {\cal N}/j}(x)$, $k=1,...,K, k\neq j$, which reflects the relative impact of each confounder among the ``observed" confounders. Therefore, treating these scores as the benchmark, we can compare the influence score for the ``unobserved" $X_j$,  $\Delta_{j, {\cal N}}(x)$, with them to study the completeness of the ``observed" confounders with index ${\cal N}/j$. In particular,  we define the following sensitivity index probability (SIP) as
$$SIP_j(x)=\frac{1}{K-1}\sum_{k=1, k\neq j}^{K} I\left(|\Delta_{j, {\cal N}}(x)|>|\Delta_{k, {\cal N}/j}(x)|\right),$$
i.e., the proportion of the observed influence scores less extreme than the true influence score for $X_j$. Clearly, if including $X_j$  in the ``observed" list does not change the propensity score much, then we expect $SIP_j(x)$ to be close to $0$. Otherwise, if $X_j$ impacts the propensity score largely, that is, $\Delta_{j, {\cal N}}(x)$ is large, $SIP_j(x)$ will be close to 1. Hence, our proposed SIP yields a quantitative measurement of the completeness of the confounders $X_k$ where $k\in {\cal N}/j$. Finally, because any future confounder can be any $X_j$, an overall SIP for evaluating the completeness of $\{X_1,...,X_K\}$ is defined as
$$SIP(x)=\frac{1}{N}\sum_{j=1}^N SIP_j(x).$$

\subsection{Implementation using observed data}

Suppose that data are collected from $n$ independent subjects whose treatment assignments and covariates are $Z_i$ and  $(X_{i1},..., X_{iK})$, respectively, for $i=1,...,n$. We wish to provide an empirical approximation to $SIP_j(x)$ for $j=1,...,K$ and $SIP(x)$. For any subset ${\cal A}\subset \{1,...,K\}$, we first compute $(\widehat\alpha_{\cal A}, \widehat\beta_{\cal A})$ by maximizing
\begin{eqnarray*}
	l_n(\alpha_{\cal A}, \beta_{\cal A})\equiv
 n^{-1}\sum_{i=1}^n\left[Z_i (\alpha_{\cal A}+X_{i,\cal A}^T\beta_{\cal A})\right.\\
\left. -\log (1+\exp\{\alpha_{\cal A}+X_{i,\cal A}^T\beta_{\cal A}\})\right],
\end{eqnarray*}
equivalently, by fitting a logistic regression of $Z$ on $X_{\cal A}$. The resulting propensity score is then
$$\widehat e_{\cal A}(x)=\widehat\alpha_{\cal A}+x^T \widehat\beta_{\cal A}.$$
In particular, we obtain $\widehat e_{\cal A}(x)$ for  the choice of ${\cal A}$ as follows:
$${\cal B}=\{1,...,K\}, \ \ {\cal B}_j=\{1,...,K\}/j, \ \ j=1,...,K$$
and
\begin{eqnarray*}
	&~&{\cal C}_j\equiv \{1,...,K\}/j, \hspace{1ex}{\cal C }_{j,k}\equiv \{1,...,K\}/\{j,k\},\\
	 &~&j, k=1,...,K, j\neq k.
\end{eqnarray*}
Correspondingly, we obtain
$$\widehat \Delta_{j, {\cal B}}(x)=\widehat e_{{\cal B}}(x)-\widehat e_{{\cal B}_j}(x)$$
and
$$\widehat \Delta_{k, {\cal C}_j}=\widehat e_{{\cal C}_j}(x)-\widehat e_{{\cal C}_{j,k}}(x).$$
Therefore, $SIP_j(x)$ can be estimated as
$$\widehat{SIP}_j(x)=\frac{1}{K-1}\sum_{k=1, k\neq j}^KI\left(|\widehat \Delta_{j, {\cal B}}(x)|>| \widehat \Delta_{k, {\cal C}_j}(x)|\right)$$
and the overall $SIP(x)$ is estimated as
$$\widehat{SIP}(x)=\frac{1}{K}\sum_{j=1}^K \widehat{SIP}_j(x).$$
Note that this estimation applies to any fixed value $x$.

To further obtain the confidence band for $\widehat{SIP}_j(x)$, $j=1,...,K$, and $\widehat{SIP}(x)$, we suggest the following resampling approach. We generate $w_1,...,w_n$ from $Exp(1)$. For each of the above index set, we estimate the coefficient of the propensity score by maximizing
$$n^{-1}\sum_{i=1}^nw_i\left[Z_i (\alpha_{\cal A}+X_{i,\cal A}^T\beta_{\cal A})-\log (1+\exp\{\alpha_{\cal A}+X_{i,\cal A}^T\beta_{\cal A}\})\right].$$
Essentially, we fit a weighted logistic regression. We then proceed the same way as before to obtain $\widehat {SIP}^w_j(x)$ , $j=1,...,K$, and $\widehat {SIP}^w(x)$. We repeat this process for many simulated $w_1,...,w_n$. The empirical distribution from this resampling approach can be used to construct the confidence interval or band for $\widehat {SIP}_j(x)$, $j=1,...,K$, and $\widehat{SIP}(x)$.

\section{Numerical Study}
\label{s:NumericStudy}

\subsection{Simulation study without unobserved confounders}

We conduct extensive simulation studies to examine the performance of the proposed SIPs. In the first simulation study, we assume that there are no unobserved confounders. More specifically, we simulate 10 observed confounding variables, denoted as $(X_1, \dots, X_{10})$, where $(X_1, \cdots, X_4)^T$ are correlated binary variables with marginal mean $(0.2, 0.6, 0.1, 0.3)^T$ and a correlation matrix
\begin{eqnarray*}
	\left({\begin{array}{cccc}
		1.00 & -0.40 & 0.00 & 0.10 \\
		-0.40 & 1.00 & 0.00 & -0.10 \\
		0.00 & 0.00 & 1.00 & 0.10 \\ 
		0.10 & -0.10 & 0.10 & 1.00 \\
\end{array}}\right)
\end{eqnarray*}
  and $(X_5, \cdots, X_{10})^T$ $\sim N_6(\bf{0}, \Sigma)$ with
\begin{eqnarray*}
	\bf{\Sigma} = \left({\begin{array}{cccccc}
		1.00 & 0.90 & 0.30 & 0.30 & 0.40 & 0.30 \\
		0.90 & 1.00 & 0.40 & 0.30 & 0.50 & 0.20 \\
		0.30 & 0.40 & 1.00 & 0.20 & 0.30 & 0.10 \\ 
		0.30 & 0.30 & 0.20 & 1.00 & 0.30 & 0.00 \\
		0.40 & 0.50 & 0.30 & 0.30 & 1.00 & 0.10 \\
		0.30 & 0.20 & 0.10 & 0.00 & 0.10 & 1.00 \\
\end{array}}\right)
\end{eqnarray*}
Treatment variable is generated from a logistic regression model with intercept 1.7 and the coefficients for $(X_1,...,X_{10})$ to be
\begin{eqnarray*}
	\bm{\beta} &=& (\beta_1, \dots, \beta_{10})^T \\
	&=& (0, 1.5, -0.5, -1.2, 4.4, -1.8, -0.3, 0, 0.9, -2)^T.
\end{eqnarray*}
We generate each dataset with sample size $500$. For each dataset, following Section 2.2, we compute $SIP_j(x), j=1,...,10$ and $SIP(x)$ for $x$ taking values from all observed covariate values. We then report the average of these indices over all $x$'s. The standard errors for the estimated SIP's are obtained from 100 resampled data.

Table \ref{table:simuRes_K10} reports the summary statistics of the SIP estimates, where the true value is obtained based on a large dataset with size 100,000,  and the coverage probability is calculated as the proportion of samples for which the true value is contained in the $95\%$ confidence intervals. Additionally, in the table,  the mean rank is the average rank of SIP's among these 10 confounders. From Table \ref{table:simuRes_K10}, the average values of the SIP estimates are close to the corresponding true values for most of confounders. The resampling-based standard errors are close to the true standard deviations and the coverage probabilities are around the nominal level. Comparing $\beta$'s and the mean ranks in Table \ref{table:simuRes_K10}, we observe that the mean ranks are generally consistent with the ranks of absolute value of beta, indicating that the SIP can characterize the importance of each variable in the propensity scores. Finally, we compare the simulated SIP estimations with the true SIPs by creating a quantile-quantile (Q-Q) plot. The error bar in the left panel of Figure \ref{figure:qqplot_comb} illustrates $95\%$ confidence interval for each SIP estimation. Because the points in the left panel of Figure \ref{figure:qqplot_comb} are around the diagonal line and the error bars are all across the diagonal line, we conclude that this set of confounders are complete for the propensity score calculation. We also note that the estimated SIPs are highly correlated with the true $\beta$'s for all the confounders.

\begin{table*}
\centering
\caption{Summary of simulation results when there are no unobserved confounders.\label{table:simuRes_K10}}

\begin{tabular*}{\textwidth}{@{}l@{\extracolsep{\fill}}r@{\extracolsep{\fill}}r@{\extracolsep{\fill}}r@{\extracolsep{\fill}}r@{\extracolsep{\fill}}r@{\extracolsep{\fill}}r@{\extracolsep{\fill}}c@{\extracolsep{\fill}}c@{\extracolsep{\fill}}c@{}}
  \hline
 & True value & Mean & True SD & Resampling SD & Coverage prob. & Mean rank & Beta \\
  \hline
SIP\_5 & 0.96067 & 0.95947 & 0.00915 & 0.00841 & 0.927 & 1.00 & 4.4 \\
  SIP\_10 & 0.73297 & 0.71097 & 0.05067 & 0.05321 & 0.968 & 3.60 & -2.0 \\
  SIP\_6 & 0.83250 & 0.83118 & 0.03396 & 0.03268 & 0.930 & 2.05 & -1.8 \\
  SIP\_2 & 0.50817 & 0.49113 & 0.08284 & 0.08191 & 0.952 & 5.62 & 1.5 \\
  SIP\_4 & 0.73937 & 0.72670 & 0.05248 & 0.05047 & 0.943 & 3.41 & -1.2 \\
  SIP\_9 & 0.36331 & 0.26380 & 0.11142 & 0.09979 & 0.778 & 8.11 & 0.9 \\
  SIP\_3 & 0.51394 & 0.47297 & 0.09200 & 0.09607 & 0.954 & 5.81 & -0.5 \\
  SIP\_7 & 0.24176 & 0.21617 & 0.12269 & 0.11245 & 0.878 & 8.49 & -0.3 \\
  SIP\_1 & 0.00051 & 0.21043 & 0.10939 & 0.10837 & 0.522 & 8.57 & 0.0 \\
  SIP\_8 & 0.11885 & 0.23574 & 0.12837 & 0.12106 & 0.821 & 8.35 & 0.0 \\
   \hline
\end{tabular*}
\vspace{5ex}
Note: The smaller mean rank is, the more important SIP is.\\
\end{table*}

\subsection{Simulation study with one unobserved confounder}

In this simulation study, we consider the situation that there is an unobserved confounder so as to aim to study the performance of SIPs. We first generate 11 covariates, for which the first 4 covariates are correlated binary variables with marginal mean $(0.2, 0.6, 0.1, 0.3)^T$ and correlation matrix
\begin{eqnarray*}
	\left({\begin{array}{cccccc}
		1.00 & -0.40 & 0.00 & 0.10 \\
  		-0.40 & 1.00 & 0.00 & -0.10 \\
  		0.00 & 0.00 & 1.00 & 0.10 \\
  		0.10 & -0.10 & 0.10 & 1.00 \\
\end{array}}\right)
\end{eqnarray*}
The other 7 covariates follow a truncated multivariate normal distribution from $-2$ to $2$ with mean $\bf{0}$ and
\begin{eqnarray*}
	\bf{\Sigma} = \left({\begin{array}{ccccccc}
		1.00 & 0.90 & 0.30 & 0.30 & 0.40 & 0.30 & 0.20 \\
  		0.90 & 1.00 & 0.40 & 0.30 & 0.50 & 0.20 & 0.50\\
  		0.30 & 0.40 & 1.00 & 0.20 & 0.30 & 0.10 & 0.00\\
  		0.30 & 0.30 & 0.20 & 1.00 & 0.30 & 0.00 & 0.30\\
  		0.40 & 0.50 & 0.30 & 0.30 & 1.00 & 0.10 & 0.70\\
  		0.30 & 0.20 & 0.10 & 0.00 & 0.10 & 1.00 & 0.10\\
  		0.20 & 0.50 & 0.00 & 0.30 & 0.70 & 0.10 & 1.00\\
\end{array}}\right) 
\end{eqnarray*}

To generate the outcome $Z$, we use a logistic regression model with intercept $0$ and the regression coefficients as
\begin{eqnarray*}
	\bm{\beta} &=& (\beta_1, \dots, \beta_{11})^T \\
	&=& (0, 0.7, -1.6, -0.8, 2, -0.6, -0.2, 0.8, 1.4, 0.3, 1.6)^T.
\end{eqnarray*}
The observed data for each subject consist of its treatment outcome and the first 10 covariate values. Thus, the last important covariate, whose coefficient is 1.6, is assumed to be unobserved. In this simulation study, each dataset has size 500 and we repeat 1000 times. We calculate the SIPs using the observed data as before and the results are summarized in Table \ref{table:simuRes_N5p5bH}. 
We note that the pattern of the SIPs are inconsistent of the true $\beta$'s. This is because when there is a significant unobserved confounder, the observed SIPs cannot fully capture the impact of all confounders.

\begin{table*}
\centering
\caption{Summary of simulation results when there is one unobserved confounders.}
\label{table:simuRes_N5p5bH}
\begin{tabular*}{\textwidth}{@{}l@{\extracolsep{\fill}}r@{\extracolsep{\fill}}r@{\extracolsep{\fill}}r@{\extracolsep{\fill}}r@{\extracolsep{\fill}}r@{\extracolsep{\fill}}r@{\extracolsep{\fill}}c@{\extracolsep{\fill}}c@{\extracolsep{\fill}}c@{}}
  \hline
 & True value & Mean & True SD & Resampling SD & Coverage prob & Mean rank & Beta \\
  \hline
SIP\_5 & 0.23184 & 0.21422 & 0.11841 & 0.11042 & 0.902 & 8.90 & 2.0 \\
  SIP\_3 & 0.36164 & 0.37370 & 0.09400 & 0.08684 & 0.905 & 7.54 & -1.6 \\
  SIP\_9 & 0.87761 & 0.87711 & 0.02493 & 0.02232 & 0.902 & 1.00 & 1.4 \\
  SIP\_4 & 0.53757 & 0.48742 & 0.12338 & 0.11117 & 0.912 & 6.16 & -0.8 \\
  SIP\_8 & 0.75497 & 0.73887 & 0.04888 & 0.04693 & 0.944 & 2.31 & 0.8 \\
  SIP\_2 & 0.42336 & 0.42570 & 0.14228 & 0.11773 & 0.841 & 6.80 & 0.7 \\
  SIP\_6 & 0.64323 & 0.62689 & 0.06131 & 0.06111 & 0.955 & 4.16 & -0.6 \\
  SIP\_10 & 0.59551 & 0.56785 & 0.08343 & 0.08222 & 0.949 & 5.10 & 0.3 \\
  SIP\_7 & 0.68051 & 0.66238 & 0.06266 & 0.06181 & 0.959 & 3.52 & -0.2 \\
  SIP\_1 & 0.00725 & 0.13389 & 0.09248 & 0.10289 & 0.835 & 9.50 & 0.0 \\
   \hline\\
\end{tabular*}
\end{table*}

\subsection{Simulation study based on real-world data}

We conduct an additional simulation study based on real-world data from a national database of 10,500 at-risk youth in the National Cross-Site Evaluation of High-Risk Youth Demonstration Grant Programs, which was funded by the Substance Abuse and Mental Health Services Administration's Center for Substance Abuse Prevention \citep{Springer2002}. This multiple-site evaluation study assessed funded prevention programs over 18 months with respect to socio-demographic risk and protective factors. For this simulation study, a sample of 547 youth whose initial uses of substance were prior to entry to the national evaluation is selected. The data set includes 213 youth in the prevention group and 334 in the comparison group. There are 22 confounding variables collected for each subject, including age, gender, race/ethnicity, family composition, family composition, family supervision, school prevention, community protection, neighborhood risk, family bonding, school bonding, self-efficacy, belief in self, self-control, social confidence, parental use attitudes, peer use attitudes, and peer use.

In the simulation study, we first standardize all covariates to obtain $\mathbf{X}$. Next, the binary outcome $Z$ is simulated in a logistic regression model with the coefficients same as the estimates from fitting a logistic regression model to the real data. We simulate 100 datasets. The average values of $\widehat{SIP}_j, j=1,2,\dots,22$ over $547$ subjects are given in Table \ref{table:simuRes_hry}, where the overall $SIP$ is estimated as the average of $\widehat{SIP}_j, j=1,2,\dots,22$. We also report in Table \ref{table:simuRes_hryID} the SIPs for a specific covariate value same as subject ID = 311031, for example. The summaries of the simulation results in both Tables \ref{table:simuRes_hry} and \ref{table:simuRes_hryID} indicate that the conclusions are similar to what we had before.

\begin{table*}
\centering
\caption{Summary of the average SIP estimates based on real-world data}
\label{table:simuRes_hry}
\begin{tabular*}{\textwidth}{@{}l@{\extracolsep{\fill}}r@{\extracolsep{\fill}}r@{\extracolsep{\fill}}r@{\extracolsep{\fill}}r@{\extracolsep{\fill}}r@{\extracolsep{\fill}}r@{\extracolsep{\fill}}c@{\extracolsep{\fill}}c@{\extracolsep{\fill}}c@{}}
  \hline
 & Mean & True SD & Resampling SD & Mean rank & Beta & Covariate \\
  \hline
SIP\_18 & 0.73903 & 0.07126 & 0.07251 & 3.36 & -0.41804 & selfcont \\
  SIP\_13 & 0.73047 & 0.07518 & 0.08553 & 3.68 & -0.35606 & neigh \\
  SIP\_21 & 0.71627 & 0.06579 & 0.08323 & 4.32 & 0.37397 & peeratt \\
  SIP\_10 & 0.69916 & 0.12321 & 0.09387 & 5.00 & 0.35250 & famsuper \\
  SIP\_22 & 0.67681 & 0.11533 & 0.11392 & 5.62 & -0.29345 & peeruse \\
  SIP\_2 & 0.62718 & 0.15947 & 0.14118 & 7.36 & -0.23163 & female \\
  SIP\_11 & 0.55194 & 0.17522 & 0.14389 & 9.85 & 0.16279 & schprev \\
  SIP\_20 & 0.54309 & 0.17187 & 0.14650 & 10.15 & -0.17966 & paratt \\
  SIP\_3 & 0.53720 & 0.06778 & 0.08389 & 10.35 & 1.02797 & indian \\
  SIP\_12 & 0.49993 & 0.18652 & 0.15612 & 11.69 & 0.13583 & compro \\
  SIP\_7 & 0.48962 & 0.08472 & 0.09357 & 12.36 & 0.71040 & white \\
  SIP\_19 & 0.45603 & 0.19266 & 0.14973 & 12.74 & -0.12684 & socconf \\
  SIP\_14 & 0.42393 & 0.18382 & 0.15696 & 13.77 & 0.10157 & fambond \\
  SIP\_9 & 0.41716 & 0.16917 & 0.16566 & 14.24 & 0.07860 & mother \\
  SIP\_1 & 0.39763 & 0.19741 & 0.15285 & 14.54 & -0.09386 & age \\
  SIP\_16 & 0.38092 & 0.19242 & 0.15497 & 14.98 & -0.19254 & selfeff \\
  SIP\_8 & 0.37717 & 0.18713 & 0.17492 & 15.05 & 0.01040 & motfat \\
  SIP\_17 & 0.35973 & 0.15375 & 0.16310 & 15.67 & 0.01291 & belself \\
  SIP\_4 & 0.36411 & 0.12213 & 0.10635 & 16.22 & 0.36481 & asian \\
  SIP\_15 & 0.33417 & 0.16197 & 0.15738 & 16.52 & 0.05921 & schbond \\
  SIP\_6 & 0.31219 & 0.13022 & 0.10622 & 17.42 & 0.44764 & hispanic \\
  SIP\_5 & 0.29660 & 0.08902 & 0.09720 & 18.12 & 0.47227 & black \\
   \hline\\
\end{tabular*}
\end{table*}

\begin{table*}
\centering
\caption{Summary of the SIP estimates for subject ID = 311031 based on real-world data}
\label{table:simuRes_hryID}
\begin{tabular*}{\textwidth}{@{}l@{\extracolsep{\fill}}r@{\extracolsep{\fill}}r@{\extracolsep{\fill}}r@{\extracolsep{\fill}}r@{\extracolsep{\fill}}r@{\extracolsep{\fill}}r@{\extracolsep{\fill}}c@{\extracolsep{\fill}}c@{\extracolsep{\fill}}c@{}}
  \hline
 & Mean & True SD & Resampling SD & Mean rank & Beta & Covariate \\
  \hline
SIP\_21 & 0.98238 & 0.03926 & 0.08080 & 1.70 & 0.37397 & peeratt \\
  SIP\_22 & 0.93619 & 0.11786 & 0.12736 & 2.69 & -0.29345 & peeruse \\
  SIP\_2 & 0.79333 & 0.17496 & 0.16902 & 5.42 & -0.23163 & female \\
  SIP\_20 & 0.76000 & 0.19997 & 0.17510 & 6.18 & -0.17966 & paratt \\
  SIP\_11 & 0.75095 & 0.22262 & 0.17958 & 6.28 & 0.16279 & schprev \\
  SIP\_10 & 0.74000 & 0.14711 & 0.14941 & 6.54 & 0.35250 & famsuper \\
  SIP\_13 & 0.70810 & 0.10235 & 0.14993 & 6.96 & -0.35606 & neigh \\
  SIP\_19 & 0.65238 & 0.24039 & 0.19705 & 8.21 & -0.12684 & socconf \\
  SIP\_5 & 0.57476 & 0.16148 & 0.16418 & 9.30 & 0.47227 & black \\
  SIP\_9 & 0.60190 & 0.20423 & 0.20846 & 9.35 & 0.07860 & mother \\
  SIP\_18 & 0.51952 & 0.11468 & 0.20603 & 10.75 & -0.41804 & selfcont \\
  SIP\_3 & 0.41952 & 0.10235 & 0.18987 & 12.71 & 1.02797 & indian \\
  SIP\_8 & 0.40857 & 0.19490 & 0.19889 & 13.37 & 0.01040 & motfat \\
  SIP\_14 & 0.36524 & 0.17831 & 0.17971 & 14.26 & 0.10157 & fambond \\
  SIP\_7 & 0.34381 & 0.09490 & 0.16808 & 14.72 & 0.71040 & white \\
  SIP\_1 & 0.33952 & 0.17736 & 0.18359 & 14.78 & -0.09386 & age \\
  SIP\_17 & 0.25857 & 0.13957 & 0.17997 & 16.36 & 0.01291 & belself \\
  SIP\_12 & 0.25714 & 0.13887 & 0.16293 & 16.71 & 0.13583 & compro \\
  SIP\_16 & 0.19762 & 0.15079 & 0.17652 & 17.56 & -0.19254 & selfeff \\
  SIP\_4 & 0.18810 & 0.09215 & 0.14152 & 18.30 & 0.36481 & asian \\
  SIP\_6 & 0.12000 & 0.08107 & 0.12176 & 19.68 & 0.44764 & hispanic \\
  SIP\_15 & 0.03095 & 0.04898 & 0.14000 & 21.20 & 0.05921 & schbond \\
  SIP & 0.49766 & 0.00814 & 0.03611 &  &  & overall \\
   \hline\\
\end{tabular*}
\end{table*}

\begin{figure*}[h]\centering
  \includegraphics[width=2in]{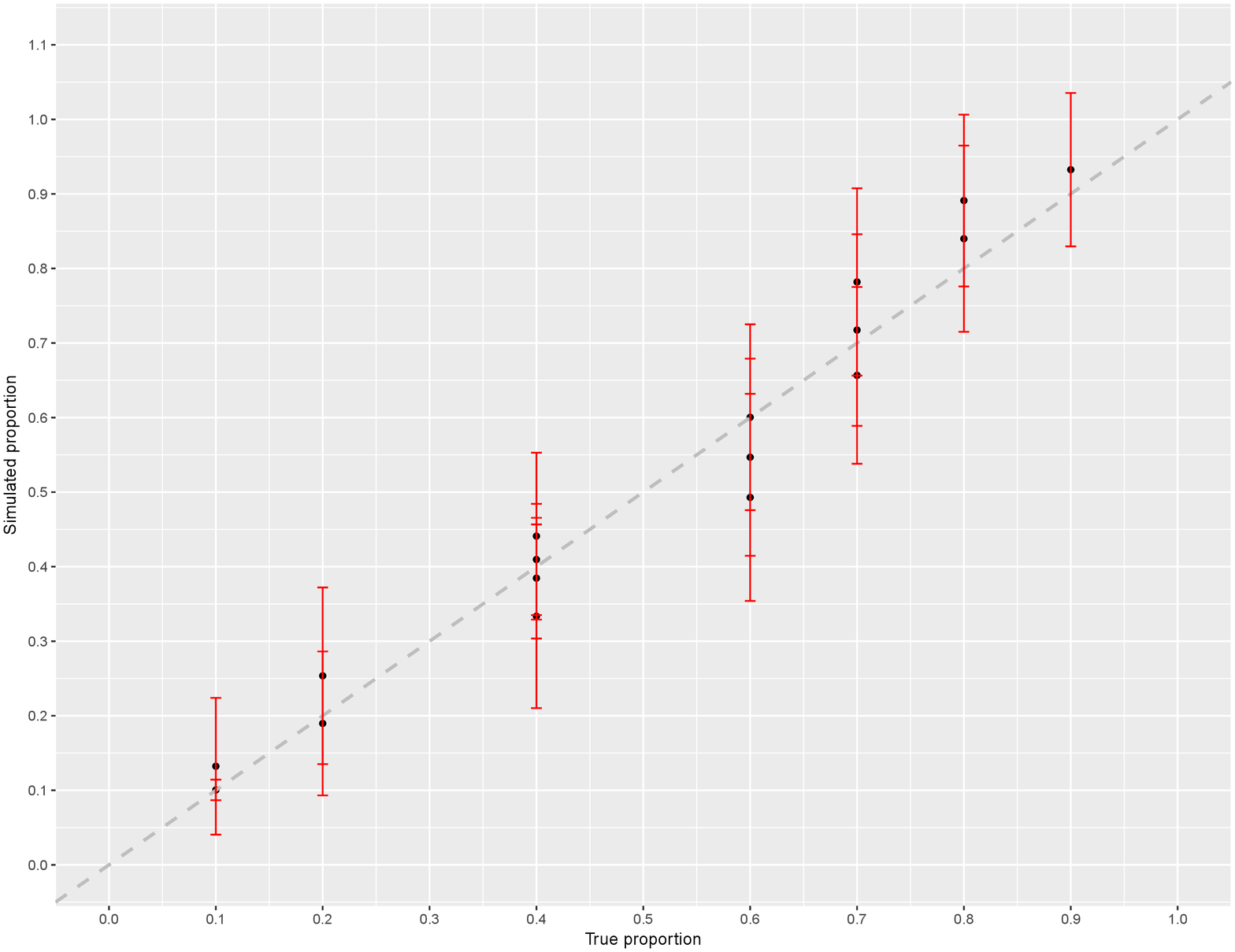} 
  \hspace{2ex}
  \includegraphics[width=2in]{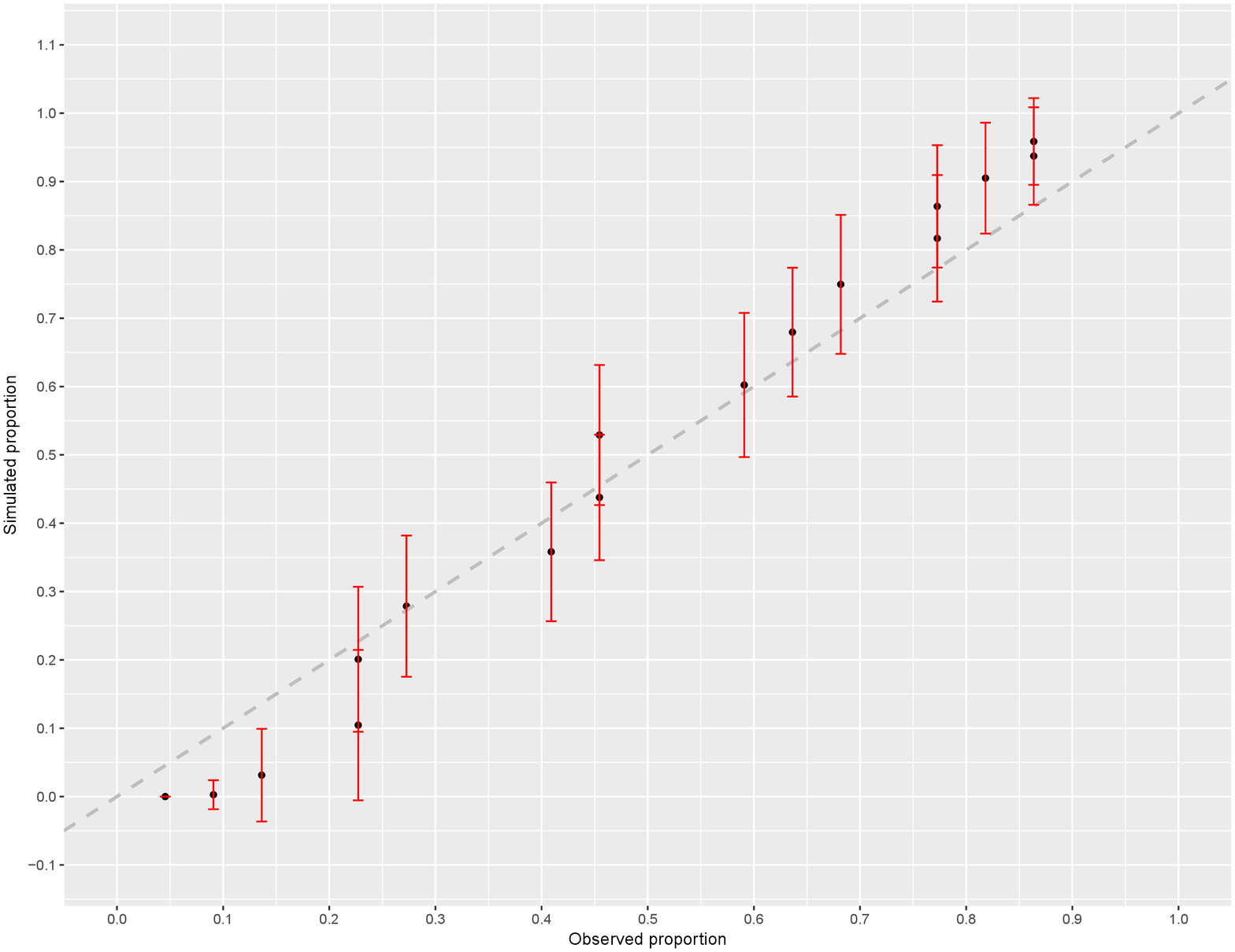} 
  \hspace{2ex}
  \includegraphics[width=2in]{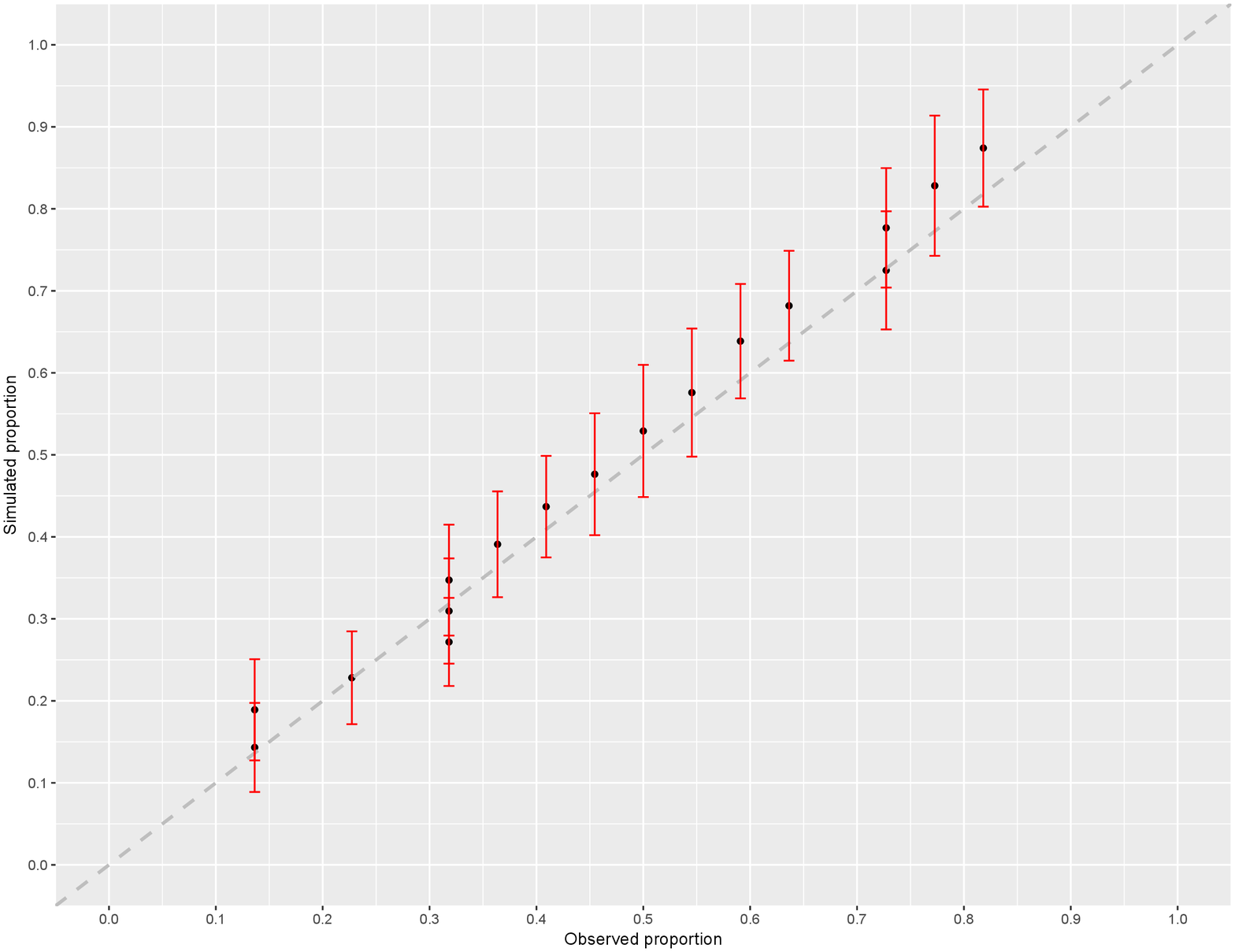} 
  \\
  Note: Left panel: Q-Q plot for the simulation study without unobserved confounders. Middle panel: Q-Q plot for the average SIP estimates based on real-world data. Right panel: Q-Q plot for the SIP estimates for subject ID = 311031 based on real-world data.
  \caption{Q-Q plots for simulation studies}
  \label{figure:qqplot_comb}
\end{figure*}

We use the SIP results from the above simulation study, for which we know there is no unobserved confounder, to compare the SIPs calculated using the real data. The results for the SIPs averaged over all subjects and for the SIPs for one particular covariate value are plotted in the middle and right panel plots of Figure \ref{figure:qqplot_comb}, respectively. The error bars in both figures indicate $95\%$ confidence interval for each SIP estimation. In the middle panel of Figure \ref{figure:qqplot_comb}, some points in bottom-left corner are below the diagonal line and error bars are not across the diagonal line, suggesting that these 22 confounders may not be sufficient for the propensity score estimation; however, for the particular covariate value from Subject 311031, the right panel of Figure \ref{figure:qqplot_comb} shows that observed confounders are sufficient in the propensity score estimation for this particular subject.

\section{Conclusion}

Unobserved confounding is a long-standing issue in causal inference using propensity score methods. Extended from the prior research on this topic, we have proposed nonparametric indices to quantify the impact of unobserved confounders through pseudo-experiments. The numerical studies suggest that the proposed indices can reflect the true impact of these confounders. The application to the real-world data provides a way to identify sufficiency or insufficiency of the observed confounders.

The pseudo-experiments are based on the assumption that the unobserved confounder is similar to some of the observed ones. This assumption can be restrictive in practice. One possible improvement is to conduct sensitivity analysis by simulating some different confounders to study their impact. It will also be interesting to develop test statistics based on the proposed indices. One possibility is to compare the derived ones vs the ones from the null. Another caveat is that although the proposed indices are robust, they may not be sensitive to detect potential confounders. One possible modification is to build some parametric or semiparametric models of the empirical indices and then evaluate the extremity under those models.

In sum, this study is one step forward to address the unobserved confounding issue in causal inference. Despite a few caveats and limitations, it is hoped that this study will stimulate further discussion on this critical issue and help move the science of causal inference forward.


\backmatter





%
  \bibliographystyle{biom} 
 \bibliography{Ref}






%



\label{lastpage}
\end{document}